\documentclass[nonacm, sigconf]{acmart}
\AtBeginDocument{%
  \providecommand\BibTeX{{%
    \normalfont B\kern-0.5em{\scshape i\kern-0.25em b}\kern-0.8em\TeX}}}

\setcopyright{acmcopyright}
\copyrightyear{2018}
\acmYear{2018}
\acmDOI{10.1145/1122445.1122456}

\acmConference[Woodstock '18]{Woodstock '18: ACM Symposium on Neural
  Gaze Detection}{June 03--05, 2018}{Woodstock, NY}
\acmBooktitle{Woodstock '18: ACM Symposium on Neural Gaze Detection,
  June 03--05, 2018, Woodstock, NY}
\acmPrice{15.00}
\acmISBN{978-1-4503-XXXX-X/18/06}

\begin{document}

\title{FiNCAT: Financial Numeral Claim Analysis Tool}


% 1st. author
\author{Sohom Ghosh}
\affiliation{%
   \institution{Fidelity Investments}
   \city{Bengaluru}
   \state{Karnataka}
   \country{India}}
\email{sohom1ghosh@gmail.com}
\orcid{0000-0002-4113-0958}

% 2nd. author
\author{Sudip Kumar Naskar}
\affiliation{%
   \institution{Jadavpur University}
   \city{Kolkata}
   \state{West Bengal}
   \country{India}}
\email{sudip.naskar@gmail.com}
\orcid{0000-0003-1588-4665}

\renewcommand{\shortauthors}{Sohom Ghosh, et al.}

\begin{abstract}
  While making investment decisions by reading financial documents, investors need to differentiate between in-claim and out-of-claim numerals. In this paper, we present a tool which does it automatically. It extracts context embeddings of the numerals using one of the transformer based pre-trained language model called BERT. After this, it uses a Logistic Regression based model to detect whether the numerals is in-claim or out-of-claim. We use FinNum-3 (English) dataset to train our model. After conducting rigorous experiments we achieve a Macro F1 score of 0.8223 on the validation set. We have open-sourced this tool and it can be accessed from \url{https://github.com/sohomghosh/FiNCAT_Financial_Numeral_Claim_Analysis_Tool}
\end{abstract}




\maketitle

\section{Introduction}
% \begin{figure}
%   \includegraphics[width=0.5\textwidth]{FiNCAT_intro.png}
%   \caption{In-claim and out-of-claim numerals in Financial Texts}
%   \label{fig:intro}
% \end{figure}
Call transcripts, financial documents relating to stocks, funds and organizations enable investors to make data-driven investment decisions. However, to persuade investors, narratives present in such documents may be just claims and not actual facts. %We illustrate such an instance in Figure \ref{fig:intro}.
Chen et. al released the NumClaim (Chinese) \cite{numclaim} and the NTCIR-16 FinNum-3 (English) \cite{finum3} datasets which comprised numerals present in financial texts and along with the annotated labels (in-claim or out-of-claim). We use the English dataset \cite{finum3} to develop \textbf{\texttt{FiNCAT}} - a tool to analyse numerals present in financial texts.

\subsection*{Our contributions}
\begin{itemize}
    \item We develop a tool to automatically detect whether numerals present in financial texts are in-claim or out-of-claim. To the best of our knowledge, we are the first one to develop such a tool.
    \item We have open-sourced\footnote{\url{https://github.com/sohomghosh/FiNCAT_Financial_Numeral_Claim_Analysis_Tool}} this tool as well as the embeddings and labels for further developments.
\end{itemize}
 %1) We develop a tool to automatically detect whether numerals present in financial texts are in-claim or out-of-claim.\\
 %2) We have open-sourced\footnote{\url{https://github.com/sohomghosh/FiNCAT_Financial_Numeral_Claim_Analysis_Tool}} this tool as well as the embeddings and labels for further developments

\begin{figure}
  \includegraphics[width=0.5\textwidth]{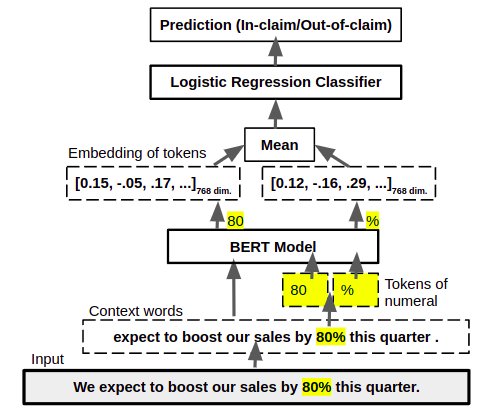}
  \caption{System Diagram of \texttt{\textbf{FiNCAT}}}
  \label{fig:system}
\end{figure}

\section{Experiments and Results}
\label{sec:exp}

\begin{table}
\centering
\caption{Model Performance on Training and Validation sets (LR=Logistic Regression, RF=Random Forest, GBM=Gradient Boosting Machine, LGBM=LightGBM, XGB=XG-Boost)}
\label{tab:results}
\begin{tabular}{lrrrr}
\toprule
       & \multicolumn{2}{c}{\textbf{Training}}                              & \multicolumn{2}{c}{\textbf{Validation}}                              \\
 \textbf{Model} & \multicolumn{1}{l}{F1-Micro} & \multicolumn{1}{l}{F1-Macro} & \multicolumn{1}{l}{F1-Micro} & \multicolumn{1}{l}{F1-Macro} \\
\midrule
 \textbf{BERT + LR}          & 0.9698          & 0.9283          & \textbf{0.9295} & \textbf{0.8223} \\
 BERT + RF          & 0.9922          & 0.9826          & 0.9211          & 0.7869          \\
 BERT + GBM         & \textbf{0.9996} & 0.9992          & 0.9270          & 0.7738          \\
 BERT + LGBM    & \textbf{0.9996} & 0.9992          & 0.9286          & 0.8009          \\
 BERT + XGB    & \textbf{0.9996} & 0.9992          & \textbf{0.9295} & 0.8054          \\
 RoBERTa + LR       & 0.9478          & 0.8694          & 0.9261          & 0.8034          \\
 RoBERTa + RF       & 0.9681          & 0.9318          & 0.8992          & 0.7461          \\
 RoBERTa + GBM      & \textbf{0.9996} & \textbf{0.9992} & 0.9219          & 0.7248          \\
 RoBERTa + LGBM & \textbf{0.9996} & \textbf{0.9992} & 0.9270          & 0.7699          \\
 RoBERTa + XGB & 0.9993          & 0.9983          & 0.9244          & 0.7588 \\
\bottomrule
\end{tabular}
\end{table}

\begin{figure*}
  \includegraphics[width=\textwidth]{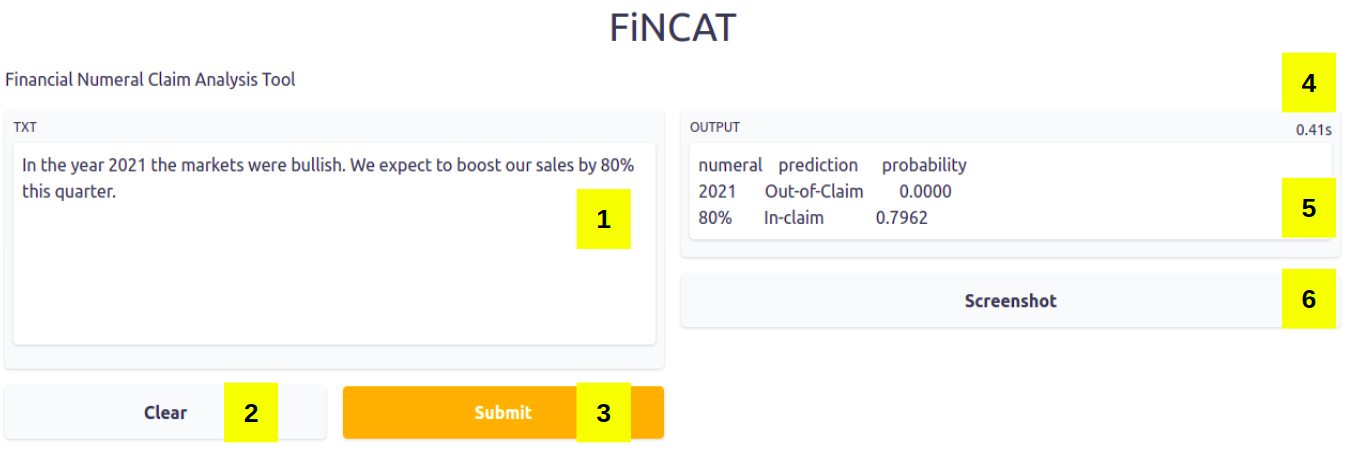}
  \caption{\texttt{\textbf{FiNCAT}}: Financial Numeral Claim Analysis Tool}
  \label{fig:tool}
\end{figure*}

We initiated by exploring the ``NTCIR-16 FinNum-3 (English): Investor’s and Manager’s Fine-grained  Claim  Detection" dataset \cite{finum3}. The training and validation set had 8,337 and 1,191 records respectively. Furthermore, each of the target numerals was labelled as in-claim or out-of-claim by experts.  Most of these financial texts had more than one target numeral. In order to deal with this, we tried to define a context window around the target numeral by considering a certain number of words before and after it. We empirically decided to use 6 words before and after the target numeral as the context window. 

We primarily experimented with two kinds of embeddings BERT-base \cite{devlin-etal-2019-bert} and RoBERTa-large \cite{Liu2019RoBERTaAR}. We extracted the mean of the embeddings of the constituent tokens of the target numeral given the words in the context window. We trained several machine learning models using the mean embeddings as features to detect whether the target numeral was in-claim or not. These models include Logistic Regression, Random Forest \cite{randomforest}, Gradient Boosting Machine \cite{gbm}, LightGBM \cite{ke2017lightgbm} and XG-Boost \cite{xgboost}. Keeping the threshold at 0.5 and we used F1 score for evaluation.

Analysing the results presented in Table \ref{tab:results}, we finally decided to move ahead with the logistic regression based model trained using BERT \cite{devlin-etal-2019-bert} embeddings (768 dimensions). It performed the best and was more efficient, explainable than the others. We present the final architecture in Figure \ref{fig:system}.

%[FINAL MODEL: SAY DELIBERATELY CHOOSE LOGISTIC REGRESSION AS IT IS PERFORMING THE BEST, LIGHT WEIGHT, NEEDS LEAST TIME TO GENERATE PREDICTIONS AND MORE EXPLAINABLE THAN THE OTHERS]

\section{Tool Description}
We deploy the tool using gradio\footnote{https://gradio.app/} on Google Colab\footnote{https://colab.research.google.com/}. We present a screenshot of it in Figure \ref{fig:tool}. It comprises six parts: 1) \texttt{\textbf{input text box}}, 2) \texttt{\textbf{clear button}}, 3) \texttt{\textbf{submit button}}, 4) \texttt{\textbf{execution time}} , 5) \texttt{\textbf{output text}} and 6) \texttt{\textbf{screenshot button}} . The \texttt{\textbf{input text box}} takes any text as input. However, since this tool is specifically built for the financial domain, we recommend users provide texts related to finance like financial conversations, annual reports of organizations and so on. On pressing the \texttt{\textbf{submit button}} we look for words in the input text which contains at least one digit. For each such word, we evaluate the model described in section \ref{sec:exp}. This consists of computing the mean of contextual BERT \cite{devlin-etal-2019-bert} embeddings of the constituent tokens present in the target numeral. This mean (768 dimensions) is used as features to score the Logistic Regression model. Finally, we generate a \texttt{\textbf{output table}} which consists of three columns: i) numerals present in the input text ii) prediction stating whether the numerals are in-claim or out-of-claim and iii) probability predicted for each of them. The \texttt{\textbf{screenshot button}} and the \texttt{\textbf{clear button}} allow users to take screenshots and clear the entered texts respectively.

We use Google Colab (free version CPU) to assess if it can detect in-claim numerals in real-time. We observe that the average time needed to generate predictions (\texttt{\textbf{execution time}}) for a given financial text consisting of 18 words and having 2 numerals is 0.25 seconds.

\section{Conclusion}
In this paper, we present a tool \texttt{\textbf{FiNCAT}} which uses context-based embeddings and machine learning to detect in-claim numerals present in financial texts. Presently, it takes only texts as input and checks for all the numerals present in the given text.

In future, we want to take the target numeral as an input from the user. This is supposed to reduce the computational time. Further tuning of hyper-parameters of the tree-based models and threshold used for prediction may yield better results. Based on the popularity we shall consider hosting it permanently using Hugging Face Spaces\footnote{https://huggingface.co/spaces}. Another interesting direction for future research would be to explore different methods for generating embeddings of the target numerals as a whole rather than taking the mean of embeddings of its constituent tokens.

%%%%%%%%\balance

\bibliographystyle{ACM-Reference-Format}
\bibliography{sample-base}

\appendix

\end{document}